\documentclass[pss]{wiley2sp}
\usepackage{graphicx}
\usepackage{amsmath}
\usepackage{bm}
\usepackage{amssymb}
\usepackage{soul}
\usepackage{amsfonts}
\usepackage{hyperref}
\DeclareMathAlphabet\mathbfcal{OMS}{cmsy}{b}{n}

\makeatother

\begin{document}

\title{Steering Magnetic Skyrmions with Nonequilibrium Green's Functions}

\author{%
Emil Vi\~nas Bostr\"om~\textsuperscript{\textsf{\bfseries 1},\Ast} and Claudio Verdozzi~\textsuperscript{\textsf{\bfseries 1},\Ast}}
\mail{%
\textsf{emil.vinas\_bostrom@teorfys.lu.se},~\textsf{claudio.verdozzi@teorfys.lu.se}}
\institute{%
\textsuperscript{1}\,Department of Physics and European Theoretical Spectroscopy Facility, Lund University, Box 118, S-22100 Lund, Sweden}

\abstract{\bf%
Magnetic skyrmions, topologically protected vortex-like configurations in spin textures, are of wide 
conceptual and practical appeal for quantum information technologies, notably in relation to the making
of so-called race-track memory devices. Skyrmions can be created, steered and destroyed with magnetic fields and/or 
(spin) currents. Here we focus on the latter mechanism, analyzed via a microscopic treatment of the skyrmion-current interaction. The system we consider is an isolated skyrmion in a square-lattice cluster, interacting with electrons spins in a current-carrying quantum wire. For the theoretical description, we employ a quantum formulation of spin-dependent currents via nonequilibrium Green's functions (NEGF) within the generalized Kadanoff-Baym ansatz (GKBA). This is combined with a treatment of skyrmions based on classical localized spins, with the skyrmion motion described via Ehrenfest dynamics. With our mixed quantum-classical scheme,
we assess how time-dependent currents can affect the skyrmion dynamics, and how this in turn depends on electron-electron and spin-orbit interactions in the wire. Our study shows the usefulness of a quantum-classical treatment of skyrmion steering via currents, as a way for example to validate/extract an effective, classical-only, description of skyrmion dynamics from a microscopic quantum modeling of the skyrmion-current interaction.}

\maketitle   


\section{Introduction}\label{Intro}
Magnetic phenomena are inherently quantal in nature~\cite{Ashcroft76,Mattis06}. Yet, in a large number of situations, the use of classical spin models to address magnetic properties of solids has successfully allowed a classification of ordinary magnetic phases such as e.g. ferro- and antiferromagnetic order~\cite{Watson70,Takahashi87,Seabra11}. The reason is to be ascribed to the fact that, conceptually, the classical picture rigorously emerges in the large-spin limit. With progress in experimental characterization, novel and more complex types of magnetic ordering have been unraveled~\cite{Ishikawa76,Pfleiderer04,Uchida08,Grigoriev09}. Still, a classical description has often provided a valuable perspective also in these more challenging instances.

A case in point is chiral magnets, where space inversion symmetry is broken by the crystal structure or a material interface, thus producing additional types of magnetic patterns~\cite{Muhlbauer09}. The situation relevant to the present work is magnetic skyrmions. These are vortex-like magnetic configurations, which can arise as a spontaneously formed hexagonal lattice in the ground state of chiral magnets, or as quasi-particle excitations on top of a ferromagnetic ground state~\cite{Skyrme62,Rossler06}. The energetic stability of skyr-mions is a consequence of an antisymmetric exchange (also called  Dzyaloshinskii-Moriya, DM) interaction~\cite{Dzyaloshinskii58}, which results from a super-exchange mechanism mediated by the spin-orbit interaction~\cite{Moriya60}. As a simple, immediate visualization for a skyrmion, one can think of a 2D ferromagnetic lattice with all spins orthonormal to the lattice plane, except for a region where spins progressively turn concentrically into an anti-parallel alignment, with the plane free energy minimized by circular
symmetry (as an example, see Fig.~\ref{fig:model}). In general, skyrmions can have either a chiral or achiral structure, termed Bloch or Ne\'el type skyrmions respectively, and can exist also in antiferromagnetic systems~\cite{Rosales15}. The type of structure that is favored is determined by the particular form of the DM interaction~\cite{Nagaosa13,Han17}, and in the present work we consider Bloch skyrmions. Further, an external magnetic field can be required to induce a stable skyrmion phase: for example, as shown in neutron scattering experiments for MnSi~\cite{Muhlbauer09}, with no magnetic field the skyrmion crystal phase is absent, and the ground state shows a helical (spiral) spin texture. 

The skyrmion configuration is interesting due to its non-trivial topological structure, that makes a skyrmion protected 
against perturbations. This feature is highly attractive from a quantum information perspective, where the topological 
charge of a skyrmion could be considered as encoding a qubit state, to be used in what has come to be known as racetrack memories~\cite{Parkin08}, where magnetic domain walls are moved via currents. It is now possible to write and delete single skyrmions using spin currents from a scanning tunneling microscope (STM) tip~\cite{Romming13,Hsu16}. However, to make use of skyrmions for quantum information and spintronic purposes, it is also necessary to manipulate skyrmions via external means~\cite{Woo16,Zhang18,Iwasaki13,Sampaio13}. 

A vast body of literature uses a classical spin framework to describe both 
static and dynamical skyrmion behavior. For example, many approaches 
to skyrmion manipulation via currents rely on time dependent numerical simulations of the 
Landau-Lifshitz-Gilbert equation~\cite{Zhang18,Iwasaki13,Sampaio13,DeLucia17}, or classical Monte Carlo steady state calculations~\cite{Rosales15,Yi09}. 
We follow the same practice here, but only as far as the spin description is concerned.
Instead, being interested in a microscopic description of how an electronic (spin) current 
affects skyrmion dynamics, we resort to a nonequilibrium quantum treatment of the electrons' spins. 
This is done in terms of a nonequilibrium Green's functions (NEGF) approach, that we merge with 
with a classical dynamics description of the skyrmion spin texture. The proposed quantum-classical
scheme is not limited to skyrmion dynamics: we are in fact able to explicitly include electronic correlations, spin-orbit effects, and general magnetic non-collinearity in the quantum subsystem, features of potential importance in several situations where classical
and quantum spins coexist in the description.

Typically, skyrmions occur on large lattice distances (on the order of tens of nanometers)~\cite{Nagaosa13}, meaning 
a large number of spin sites are involved. This implies, for our quantum-classical approach, a NEGF
treatment of fairly large samples. In addition, skyrmion and electron dynamical responses take place on rather different timescales,
so that long simulation times of the electronic Green's functions are required. These two features together
make the problem computationally very hard when using double-time NEGF formulations. 

To put our approach within practical reach,
we use a time-diagonal formulation of NEGF, based on the so-called generalized Kadanoff
Baym ansatz (GKBA)~\cite{Lipavsky86}. With few exceptions~\cite{Kalvova18}, the NEGF-GKBA has so far been
used only for spin-compensated systems~\cite{Hermanns12,Balzer13-2,Latini14,Bostrom18,Hopjan18,Perfetto18,Perfetto18-2}. In this work, we 
take into account spin-orbit effects and magnetic non-collinearity, in a theoretical framework where electrons mutually interact.

Concerning the use of the GKBA in a spin-dependent approach, it has been known for a while that (even for non-interacting and/or spin-compensated systems) the GKBA introduces an error for reservoirs with finite energy support, but approaches the correct solution in the wide band limit (WBL)~\cite{Latini14}. Recently, it was pointed out that this problem might become more severe for magnetic systems \cite{Kalvova18}, indicating the importance of including corrections beyond the standard GKBA to describe magnetic quantum-transport setups with finite leads. In our work, we consider only the WBL regime, and thus it is not necessary to consider such corrections.

With our proposed methodology, we study how (spin polarized) currents flowing in a quantum
wire can induce and influence the dynamics of an isolated classical skyrmion. We also address how electronic
correlations and spin-orbit effects in the wire affects the skyrmion behavior. 
We proceed as follows: in Sect.~\ref{SandH}, the system and the corresponding Hamiltonian are described.
The solution of the classical skyrmion problem (i.e. in the absence of currents)
is discussed in Sect.~\ref{GS}, and followed in Sect.~\ref{GKABncl} by a presentation of the 
non-collinear, spin-dependent GKBA. In Sect.~\ref{Coupled}, details of the mixed quantum-classical
scheme are provided, followed by results and their discussion in Sect. \ref{Timeevo}.
A few general remarks are given at the end, in Sect. \ref{conc}.


\section{System and Hamiltonian}\label{SandH}
\begin{figure}[t]
 \includegraphics[width=0.5\textwidth]{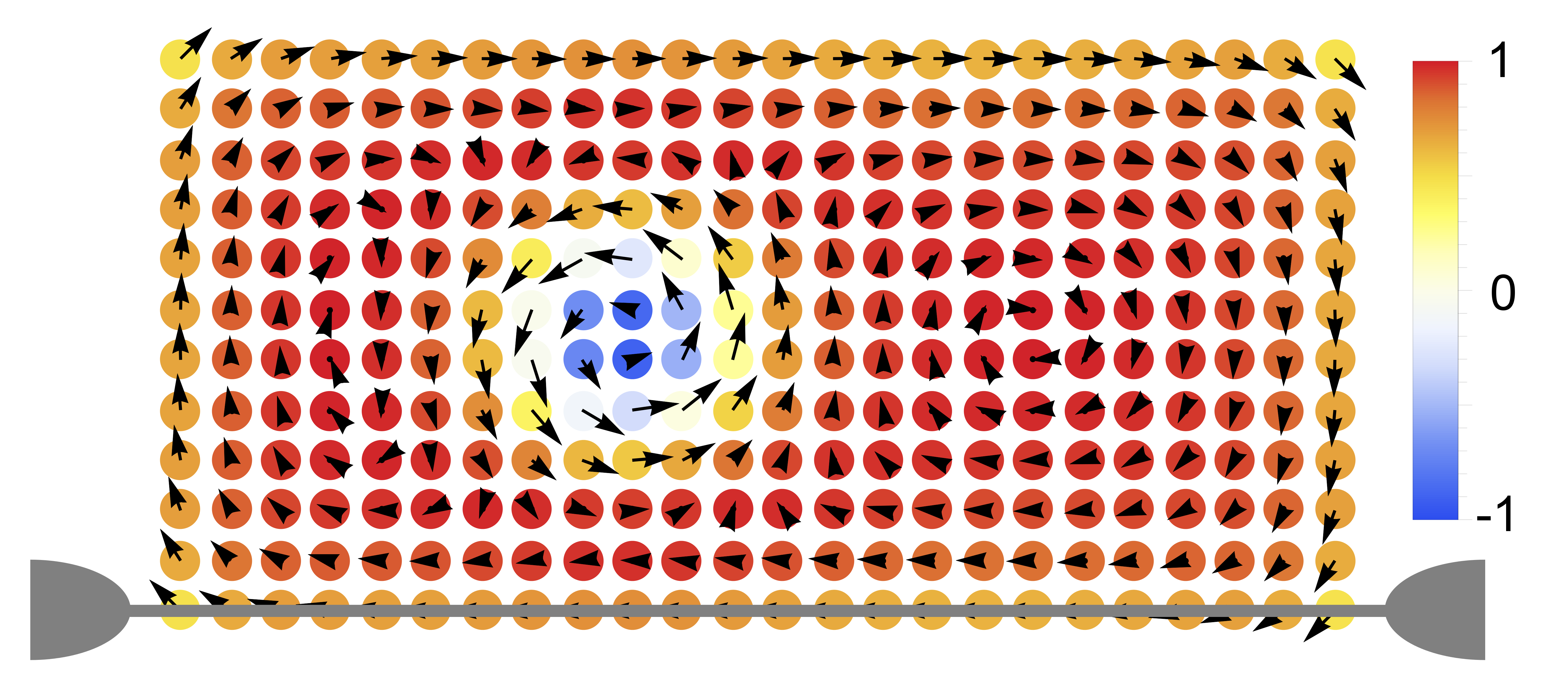}
 \caption{Schematic illustration of the system. A classical spin texture on a square lattice (in this case, representing an isolated skyrmion) interacts with the electrons in a wire carrying a (spin polarized) current. The device part of the wire and the embedding leads are shown explicitly as superimposed at the bottom. The array of colored dots represents the value of the $z$-component of the localized spins according to the vertical color bar, whilst the in-plane ($x,y$) spin-projections are shown explicitly by arrows. The color coding specifications apply to all figures in the paper.}
 \label{fig:model}
\end{figure}

We consider mutually interacting Heisenberg spins on a square lattice, which are magnetically coupled to the electronic spins of a quantum wire. The system (Fig.~\ref{fig:model}) is described by the Hamiltonian
\begin{align}
H = H_s + H_w + H_{sw}.
\end{align}
The first part of $H$ governs the dynamics of the spins:
\begin{align}
H_s &= -J\sum_{\langle mn \rangle} \hat{{\bf S}}_m\cdot \hat{{\bf S}}_n - D\sum_{\langle mn \rangle} \hat{\bf e}_{mn}\cdot (\hat{{\bf S}}_m\times\hat{{\bf S}}_n) \nonumber \\
&+ A_1\sum_m \sum_{i=x,y,z} (\hat{S}_m^i)^4 - A_2\sum_{\langle mn \rangle } [\hat{S}_m^x \hat{S}_{n}^x + \hat{S}_m^y\hat{S}_{n}^y]\nonumber \\
& - h(t)\sum_m \hat{S}_m^z. 
\end{align}\label{Hspin}
Here, $J$ and $D$, both taken positive, give the respective strengths of the exchange (in our case, ferromagnetic) and
Dzyaloshinskii-Moriya (DM) interactions \cite{Dzyaloshinskii58,Moriya60}. Also, the subscript $m$ refers to the spin at  
position ${\bf R}_m$ in the 2D lattice, and $\langle mn \rangle$ restricts the double sums to pairs of nearest-neighbor
lattices sites. The DM vector is given by $\hat{\bf{e}}_{mn}\equiv ({\bf R}_m-{\bf R}_n)/|{\bf R}_m-{\bf R}_n|$, favoring formation of Bloch-type skyrmions. The terms proportional to $A_1$ and $A_2$ are anisotropy terms appropriate for a square lattice
(we assume the lattice lies in the $xy$-plane), whose main effect is to break the degeneracy of the ground state~\cite{Han17,Yi09}. 
Finally, the last term in Eq.~\ref{Hspin} describes the Zeeman interaction with an external, in principle time dependent,
magnetic field $h(t)$, conventionally oriented along the $z$-direction. 

The second part of $H$  pertains to the quantum wire, and comprises three terms,
\begin{align}
H_w = H_c + H_{res} + H_{c-res}.
\label{Full_H} 
\end{align}
The first term, $H_c$, describes the central region of the wire (i.e. the ``device'') within a tight-binding picture:
\begin{align}\label{eq:ham_c}
H_c &= U\sum_i \hat{n}_{i\uparrow}\hat{n}_{i\downarrow} + \sum_{i\sigma\sigma'}c_{i\sigma}^\dagger(\epsilon_i I - h(t)\sigma_z)_{\sigma\sigma'}c_{i\sigma'} \\
 &+ \sum_{i\sigma\sigma'} \left[c_{i\sigma}^\dagger (tI + it_{so}\sigma_y)_{\sigma\sigma'}c_{i+1,\sigma'} + h.c.\right]. \nonumber 
\end{align}
Here, $c^\dagger_{i\sigma}$ ($c_{i\sigma}$) creates (destroys) an electron of spin projection $\sigma$ at site $i$, and
$\hat{n}_{i\sigma}= c^\dagger_{i\sigma} c_{i\sigma}$. The single-particle Hamiltonian is of the general form $h_{i\sigma j\sigma'} = h_{ij}\cdot \sigma$, where $\sigma = (I,{\bm \sigma})$, $I$ is the $2\times 2$ identity matrix, and ${\bm \sigma}$ is the vector of Pauli matrices. In the following however, we restrict to the explicit form of $h_{ij}$ given in Eq.~\ref{eq:ham_c}. Concerning the parameters in the wire Hamiltonian, 
$\epsilon_i$ is a local, in principle site-dependent, potential energy and $h$ the
external magnetic field (the same as in Eq.~\ref{Hspin}), while $t$ and $t_{so}$ respectively describe spin-preserving and 
spin-flip hoppings between the sites of the wire. In particular, $t_{so}$ accounts for spin-orbit effects within a tight-binding description. For simplicity, we consider only on-site interactions among electrons, i.e. we use interactions
of the Hubbard form, whose strength is given by the parameter $U$. The approach is however general enough to include
arbitrary two-body interactions of the form $v_{ijkl}$. 

The reservoirs (leads) are taken as one-dimensional, and are described similarly in a tight-binding picture, via
\begin{align}
H_{res} = t'\sum_{i\sigma\alpha} (a_{i\sigma\alpha}^\dagger a_{i+1,\sigma\alpha} + h.c.) + \sum_{i\sigma\alpha}u_{\alpha\sigma}(t)\hat{n}_{i\sigma\alpha},
\end{align}
where $t'$ is the hopping amplitude, $u_{\alpha\sigma}$ a time-dependent bias, and $\alpha$ a reservoir index. Throughout this work we consider spin-polarized reservoirs. The central region of the wire interacts with the reservoirs via the tunneling term
\begin{align}
H_{c-res} &= t_l\sum_{\sigma} a_{1\sigma l}^\dagger c_{1,\sigma} + t_r\sum_{\sigma} a_{1\sigma r}^\dagger c_{N,\sigma} + h.c.,
\end{align}
where $t_l$ and $t_r$ determine the hopping amplitude into the left and right leads.

Finally, localized Heisenberg spins and electron spins in the device part of the wire
mutually interact through an $s\!\!-\!\!d$-type coupling, given by the Hamiltonian term
\begin{align}\label{eq:interaction}
H_{sw} = \sum_{im} J'_{im} \hat{{\bf S}}_m\cdot \sum_{\sigma\sigma'}c_{i\sigma}^\dagger {\bm \sigma}_{\sigma\sigma'} c_{i\sigma}.
\end{align}
We note that, in general, the coupling strength $J'$ can be distance-dependent, and thus the wire
can be positioned arbitrarily within the lattice of Heisenberg spins. Here, however, we consider for simplicity a
straight wire aligned on top of a line of spins, as shown in Fig.~\ref{fig:model}. We consider the case $J'_{im} = J'\delta_{im}$, where at each site of the wire the electron spins interact only with the corresponding localized Heisenberg spin at the same lattice site. However, we have checked that our results are robust against changing the coupling, by instead taking e.g. $J'_{im} = J'\delta_{im_x}$.

A remark concerning the energy units for the different contributions to $H$: in the paper, all the parameters for $H_s$ are 
expressed in units of $J$, the exchange interactions for the localized spins. In turn, $J$ and 
all the electronic parameters in $H_w+H_{sw}$, including the $s\!\!-\!\!d$ exchange term $J'$, are expressed 
in units of $t$, the spin-preserving electronic hopping term, which is thus our basic energy unit.


\section{Ground state spin configuration}\label{GS}
In the large-spin limit (formally, when $\langle{\bf S}_m^2\rangle \to \infty$), spin fluctuations can to a good approximation 
be neglected, and the spins can be treated classically~\cite{Lieb73,Frohlich07}. 
In what follows, we i) assume that we are in this regime, and ii) 
for convenience normalize the {\it now classical} spin vectors, by taking
${\bf S}_m \rightarrow \tilde{{\bf S}}_m \equiv {\bf S}_m/|{\bf S}_m|$. This simply amounts to a rescaling of the spin
parameters in the Hamiltonian (for notational convenience, the tilde superscript is dropped henceforth).

To find the ground state of the classical spin system, we use simulated annealing together with the Monte Carlo
(MC) Metropolis algorithm to minimize the energy. Starting from a random spin configuration, the system temperature
is gradually decreased, while minimizing the energy in each step. Once the target temperature is  
reached, the MC routine is invoked an additional $n_{av}$ times: the ground state configuration is then taken as the 
algebraic average of these $n_{av}$ configurations.

\begin{figure}[t]
 \includegraphics[width=0.5\columnwidth]{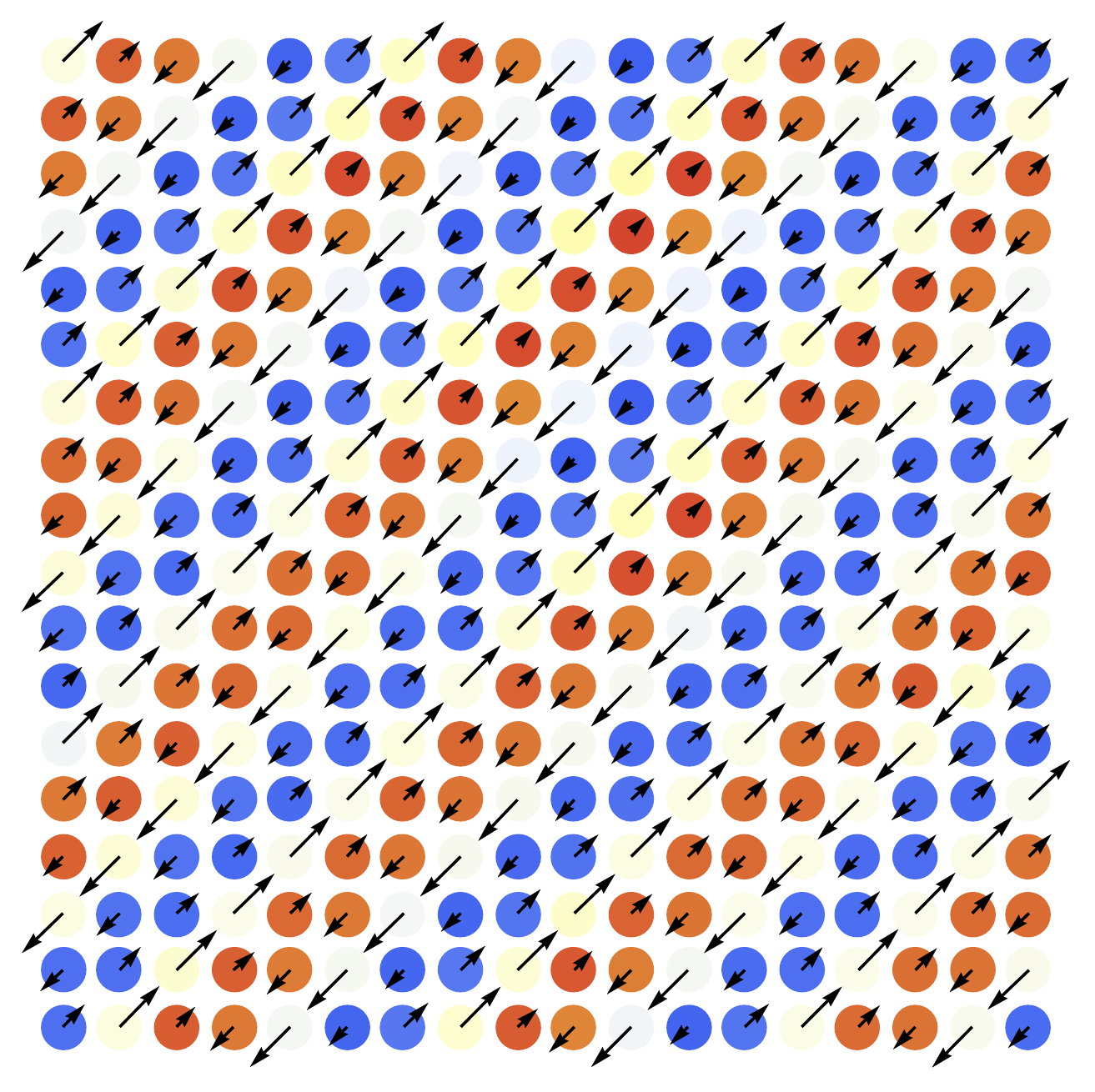}%
 \includegraphics[width=0.5\columnwidth]{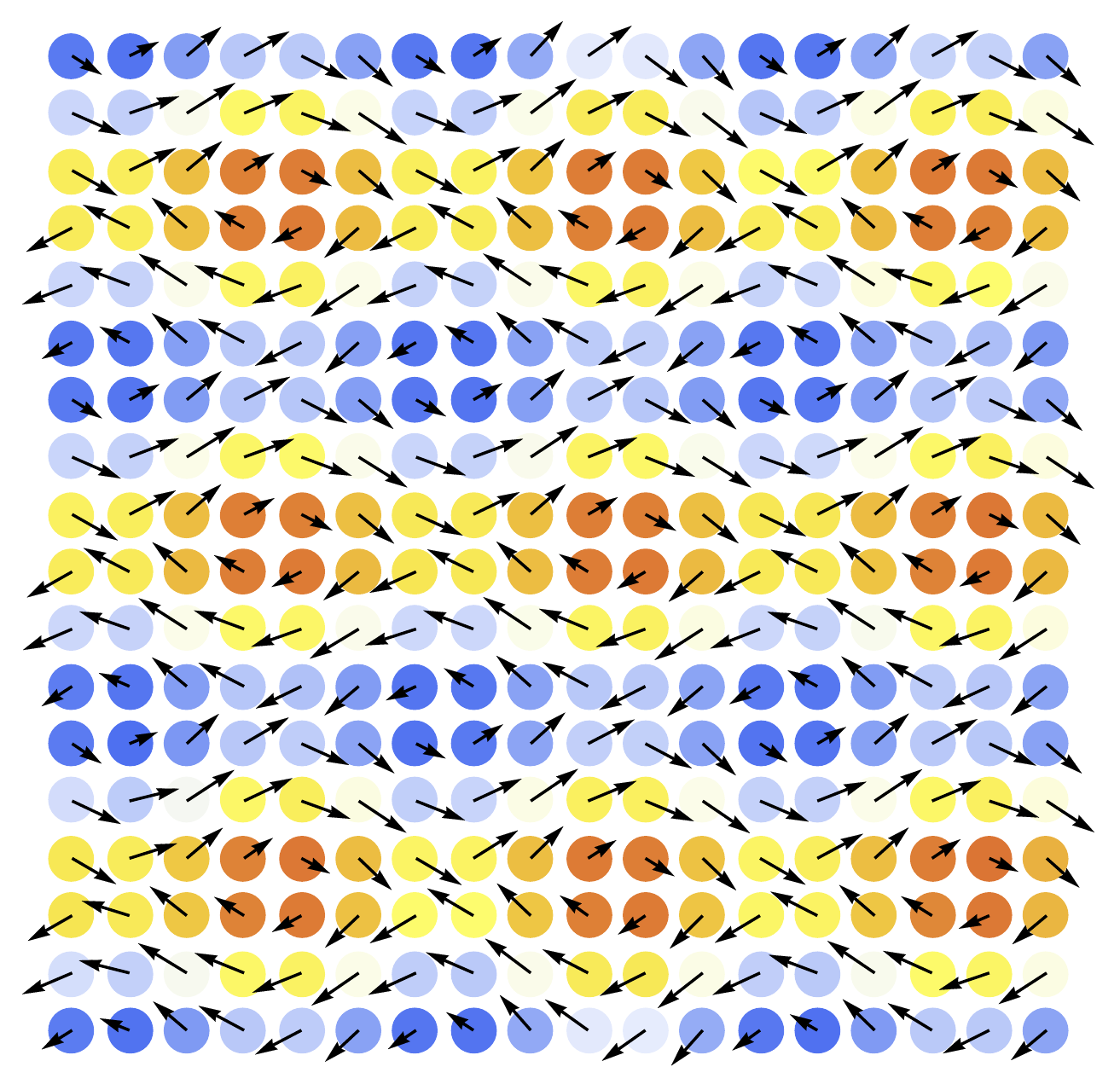}
 \includegraphics[width=0.5\columnwidth]{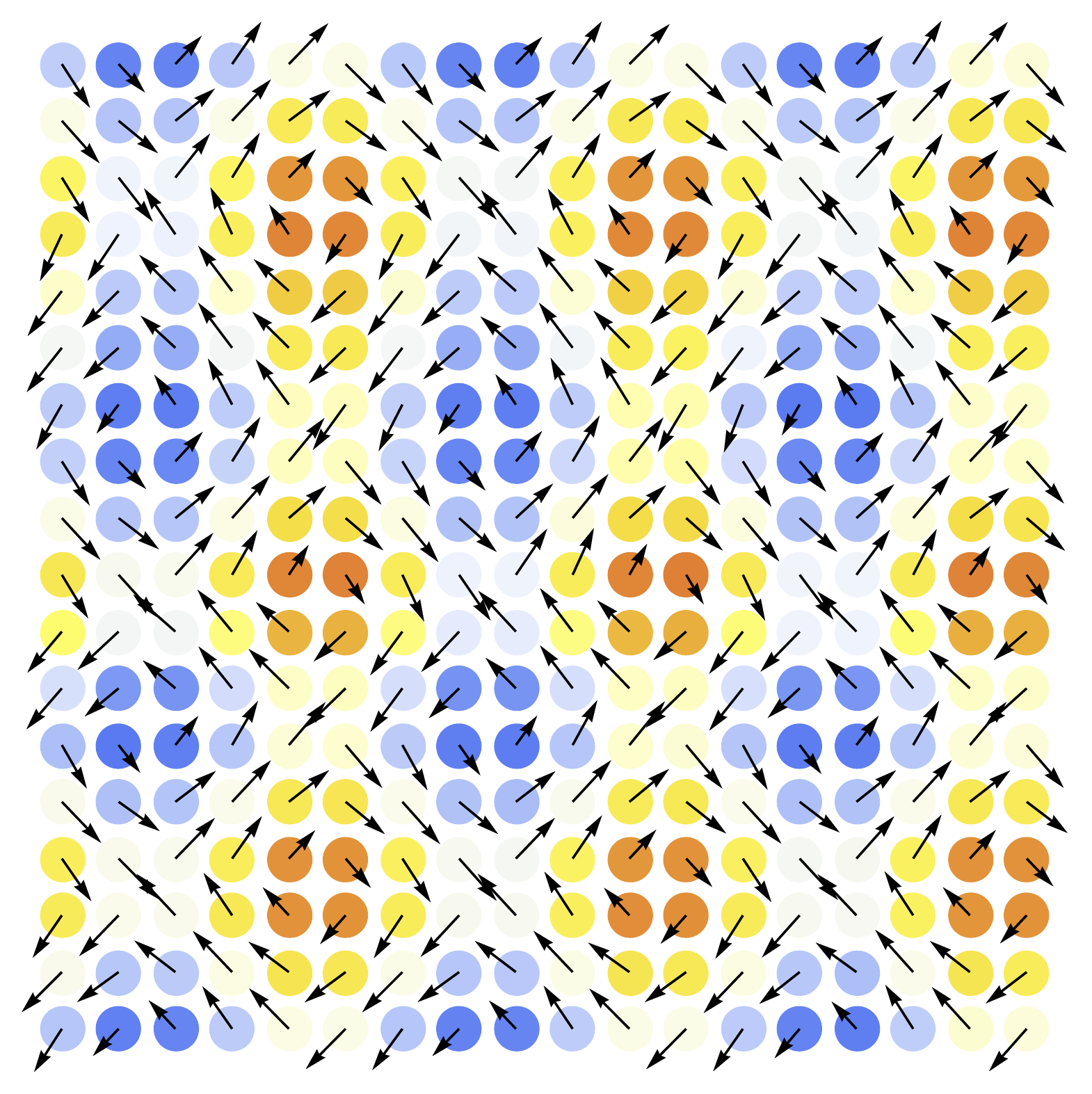}%
 \includegraphics[width=0.5\columnwidth]{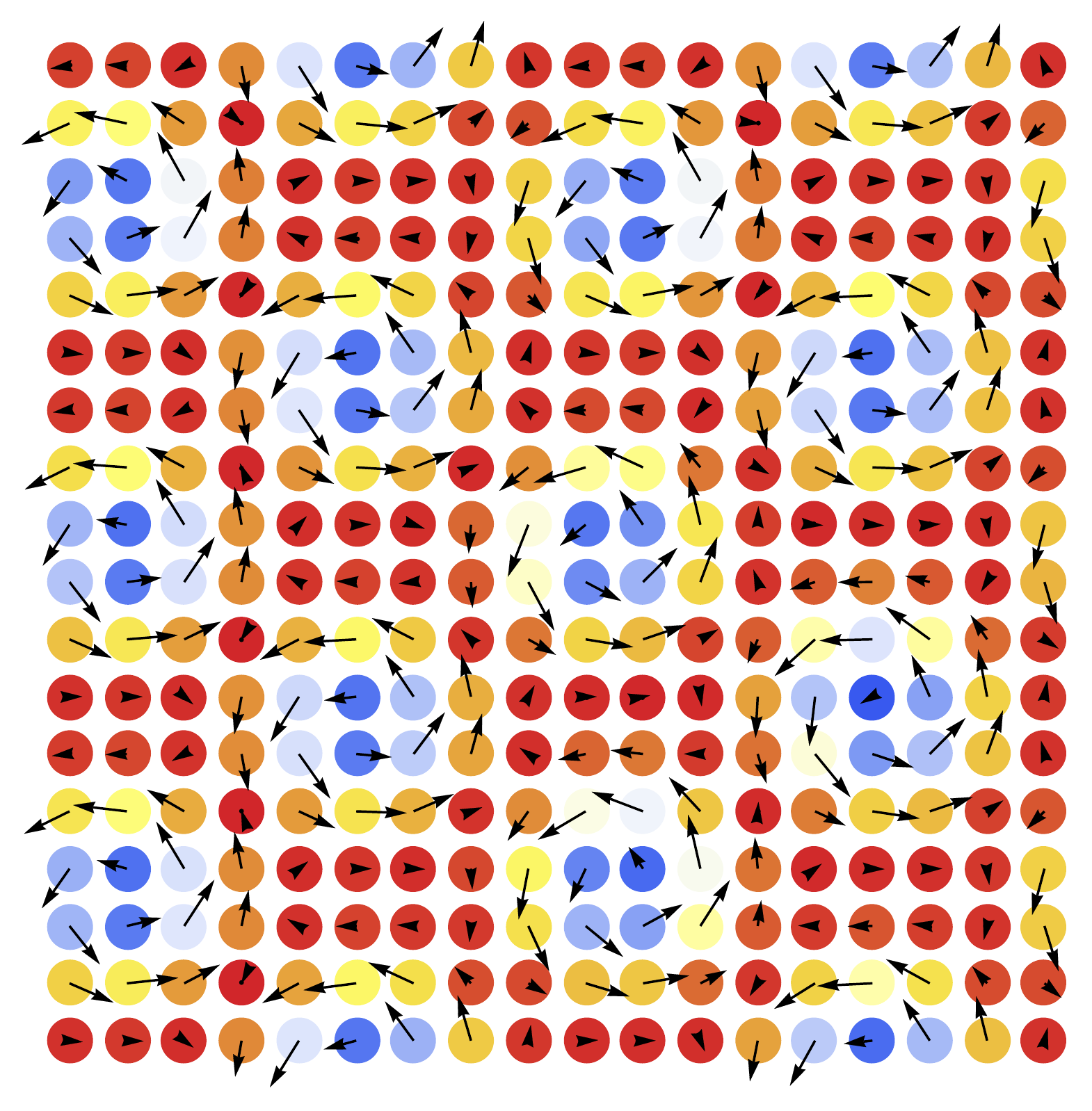}
 \caption{Spin textures obtained by varying the parameters $A_1$, $A_2$ and $h$. We take $J = 1$ to define the unit of energy, keep $D = \sqrt{6}$ fixed to have a configuration with wavelength $\lambda = 6$, and take $A_1 = 0.5$. In the top left panel $A_2 = h = 0$, while in the top right panel $A_2 = 2$, and $h = 0$. In the bottom left panel $A_2 = 3$ and $h = 0$, while in the bottom right panel $A_2 = 0$ and $h = 2$. Color coding specifications are the same as in Fig.~\ref{fig:model}.}
 \label{fig:spins}
\end{figure}

Depending on the value of the parameters $J$, $D$, $h$, $A_1$ and $A_2$, the spin Hamiltonian $H_s$ shows a rich
phase diagram (for a full discussion see~\cite{Yi09}). For zero magnetic field the ground state configuration is a so-called
spiral state, as shown in the top left panel of Fig.~\ref{fig:spins} for an $18 \times 18$ square spin lattice with periodic
boundary conditions. The sign of the parameter $D$ determines the chirality of the spiral, while the ratio $D/J$
determines the wavelength of the spiral. The results in Fig.~\ref{fig:spins} correspond to
$J = 1$ and $D = \sqrt{6}$. In addition, the propagation direction of the spiral is fixed by the sign of $A_1$: For $A_1>0$ it 
is along the $(1,1)$ direction, while for $A_1<0$ it is along the $(1,0)$ direction. 

\begin{figure}[t]
 \includegraphics[width=0.49\columnwidth]{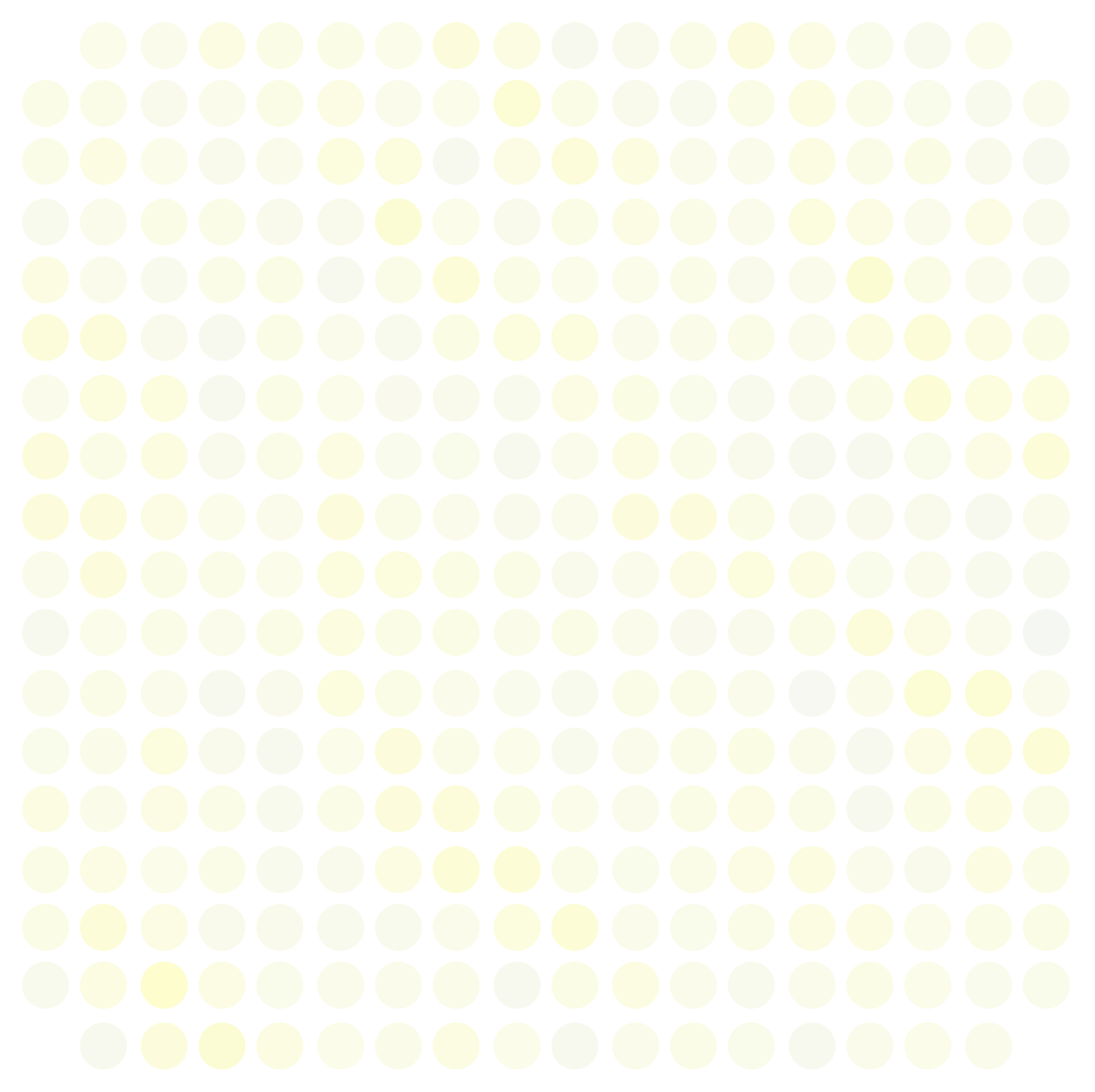}%
 \includegraphics[width=0.48\columnwidth]{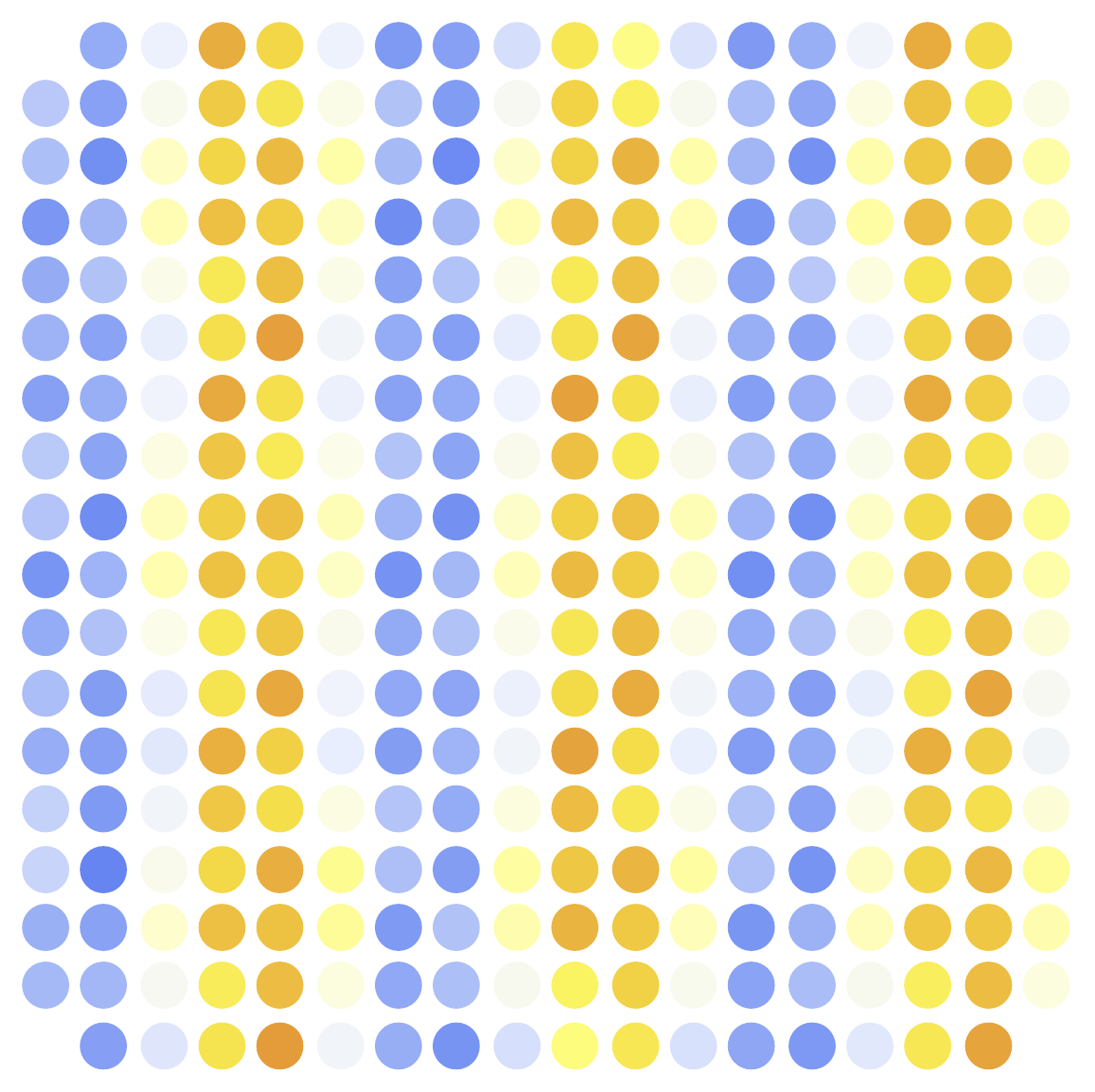}
 \includegraphics[width=0.49\columnwidth]{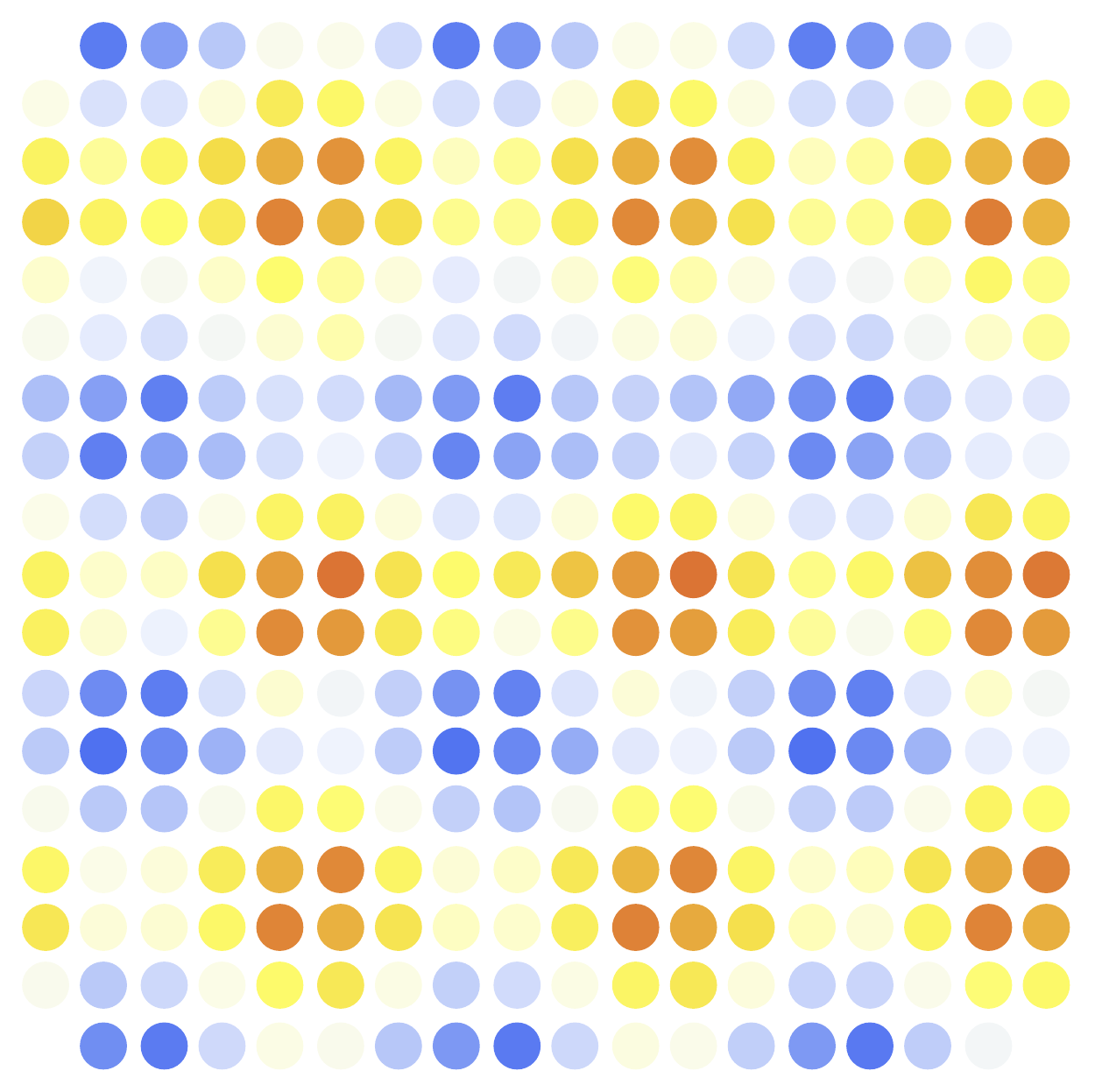}%
 \includegraphics[width=0.49\columnwidth]{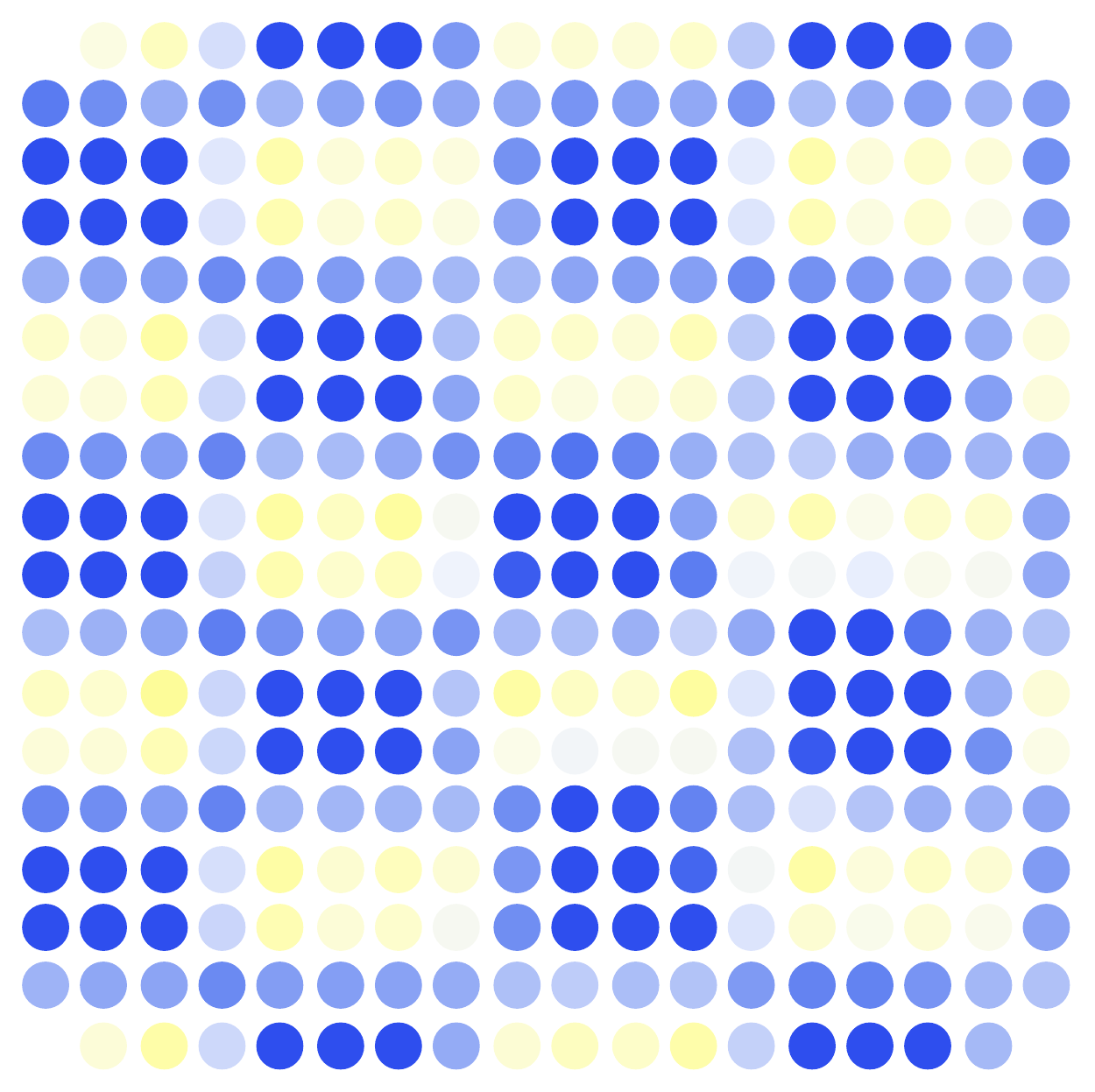}
 \caption{Skyrmion densities corresponding to the spin textures shown in Fig.~\ref{fig:spins}. Color coding specifications are the same as in Fig.~\ref{fig:model}.}
 \label{fig:densities}
\end{figure}

In Fig.~\ref{fig:densities}, we show the corresponding skyrmion density, defined as
\begin{align}
\mathbb{\varrho}^\mathcal{SK}_m = \frac{{\bf S}_m \cdot \left({\bf S}_{m+\hat{\bf x}}\times {\bf S}_{m+\hat{\bf y}} + {\bf S}_{m-\hat{\bf x}}\times {\bf S}_{m-\hat{\bf y}}\right)}{8\pi},
\end{align}
with $\hat{\bf x}$ and $\hat{\bf y}$ are the unit vectors of the square lattice. This is a compact, convenient indicator for the skyrmion content of given spin texture; for example, in the spiral phase, $\mathbb{\varrho}^\mathcal{SK}_m = 0$. When integrated over the entire plane the skyrmion density gives the topological charge $Q$, that in the continuum is conserved and takes on integer values~\cite{Nagaosa13}.

Taking $A_2 \neq 0$ gives a skyrmion crystal state, shown in the top right and bottom left panels of Fig.~\ref{fig:spins}. This state can be seen as a superposition of two spin spirals with wave vectors ${\bf k} = \pm\hat{\bf x}$ and ${\bf k} = \pm\hat{\bf y}$, and depending on if the amplitudes of the two spirals are different (equal), we find the configuration in the top right (bottom left) panel of Fig.~\ref{fig:spins}. By taking $A_2 = 0$ and $h \neq 0$ we find a skyrmion crystal that consists of three superposed spin spirals, shown in the bottom right panel of Fig.~\ref{fig:spins}. As also seen from the skyrmion densities in Fig.~\ref{fig:densities}, this state corresponds to a hexagonal lattice of skyrmions.

\section{GKBA for non-collinear spins}\label{GKABncl}
To obtain the expectation value of any time-local one-body electronic operator, it is enough to have complete 
knowledge of the time-dependent one-particle density matrix $\rho$~\cite{Balzer13-1,Stefanucci13,Hopjan14}. 
When spin-orbit effects and/or  spontaneous magnetization are present, or when external magnetic fields have a non-uniform orientation in space and time (these instances are generically referred to as a {\it non-collinear spin regime}), $\rho$ retains a full matrix
dependence on the spin indexes, i.e. $\rho \equiv \rho_{ij}^{\sigma\sigma'}(t)$. Here, with reference to 
Eqs.~\ref{Full_H}, $i$ and $j$ are site indexes, $\sigma$ and $\sigma'$ are spin indexes, and $t$ is the time variable.

In this section we provide the equation of motion for $\rho_{ij}^{\sigma\sigma'}(t)$
in the non-collinear spin case using nonequilibrium Green's functions (NEGF) within the generalized Kadanoff-Baym ansatz~\cite{Lipavsky86}. Our starting point is the contour ordered nonequilibrium Green's function $G$, defined by
\begin{align}
G_{ij}^{\sigma\sigma'}(z,z') = -i\langle\mathcal{T}\left[c_{i\sigma,H}(z)c^\dagger_{j\sigma',H}(z')\right]\rangle.
\end{align}
Here, the expectation value is taken with respect to the grand-canonical ensemble, the arguments $z$ and $z'$ of the Green's function take values on the Keldysh contour $\gamma$, going from $t = -\infty$ to $t = \infty$ and back again, and the contour-ordering operator $\mathcal{T}$ orders operators so that later contour time arguments are to the left. The creation and annihilation operators are taken in the Heisenberg picture. In the present case, the equation of motion for $G$ must retain the full spin structure for all quantities involved:
\begin{align}\label{Geq}
\bigg(i\frac{d}{dz} &-[h^{HF}(z;\mathbfcal{S} )]_{ik}^{\sigma\sigma''}\bigg)G_{kj}^{\sigma''\sigma'}(z,z') = \delta_{ij}^{\sigma\sigma'}\delta(z,z') \\
                 &+ \int_\gamma dz_1 \Sigma_{ik}^{\sigma\sigma''}(z,z_1)G_{kj}^{\sigma''\sigma'}(z_1,z'), \nonumber 
\end{align}
with a corresponding equation being satisfied by its adjoint. In Eq.~\ref{Geq} and onwards, repeated spin and site indexes are implicitly summed over. Further, $h_{HF}$ is the single-particle Hamiltonian of the system, which depends parametrically on the set of time-dependent classical spins ${\bf  \mathbfcal{S}}\equiv\{{\bf S}_1, {\bf S}_2,\ldots,{\bf S}_N\}$. In addition $h_{HF}$ contains the Hartree-Fock contribution of the interactions, while any correlation effects beyond this approximation are subsumed in the self-energy $\Sigma$. Eq.~\ref{Geq} and its adjoint can be decomposed using the Langreth rules~\cite{Langreth76} into a set of coupled equations known as the Kadanoff-Baym equations (KBE)~\cite{Kadanoff62}, where the Green's functions depend on real time arguments. 

Since the Green's function depends on two time arguments, and the integral kernel retains the full memory of the system, the solution of the KBE scales at least cubically with time \cite{Hermanns12}. 
However, since  $\rho_{ij}^{\sigma\sigma'}(t) = -iG_{ij}^{\sigma\sigma',<}(t,t')$, it is possible starting from the equation of motion for the lesser Green's function $G^<$ and its adjoint, to derive an equation of motion directly for the density matrix as
\begin{align}\label{eq:eom}
\frac{d}{dt}\rho_{ij}^{\sigma\sigma'}(t) +i\left[h_{HF}(t;\mathbfcal{S}), \rho(t)\right]_{ij}^{\sigma\sigma'}\!\!\!\!=\!-\left(I_{ij}^{\sigma\sigma',<}(t) + h.c.\right),
\end{align}
with 
\begin{align}\label{eq:coll}
I_{ij}^{\sigma\sigma',<}(t) = \int &dt' \bigg[\tilde{\Sigma}_{ik}^{\sigma\sigma'',<}(t,t')G_{kj}^{\sigma''\sigma',A}(t',t) \\
                         &+ \tilde{\Sigma}_{ik}^{\sigma\sigma'',R}(t,t')G_{kj}^{\sigma''\sigma',<}(t',t)\bigg]. \nonumber
\end{align}
The self-energy $\tilde{\Sigma}$ in Eq.~\ref{eq:coll} is made out of two parts:
in matrix notation (and with spin/site indexes not explicitly shown), $\tilde{\Sigma}=\Sigma_{emb.}+\Sigma$, where 
$\Sigma_{emb.}$ is the embedding self-energy that describes the effect of the macroscopic leads 
to which the wire is connected~\cite{Myohanen08}, and $\Sigma$ accounts for the electronic correlations in the wire.

Eq.~\ref{eq:eom} is not a closed equation for $\rho(t)$, since the collision integral $I^<(t)$ (Eq.~\ref{eq:coll}) depends on $G^<$ for times $t'\neq t$. An (approximate) closure can be made via the generalized Kadanoff-Baym ansatz (GKBA)~\cite{Lipavsky86}
\begin{align}\label{eq:GKBA}
G_{ij}^{\sigma\sigma',<}(t,t') &\approx \rho_{ik}^{\sigma\sigma''}(t)G_{kj}^{\sigma''\sigma',A}(t,t') \\
                            &- G_{ik}^{\sigma\sigma'',R}(t,t')\rho_{kj}^{\sigma''\sigma'}(t'), \nonumber
\end{align}
where the retarded Green's function $G^R(t,t')$ is assumed to be of the form 
\begin{align}\label{eq:g_ret}
G^R(t,t') = -i\theta(t-t')T e^{-i\int_{t'}^t dt_1 h_{qp}(t_1)}.
\end{align}
The quasi-particle Hamiltonian $h_{qp}$ will in the following be taken to be $h_{HF}$. 

\begin{figure}
\centering
\includegraphics[width=\columnwidth]{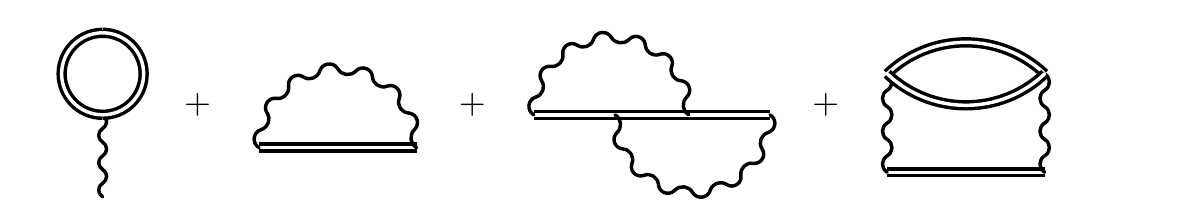}
\caption{Feynman diagrams contributing to the second Born self-energy of Eq.~\ref{eq:sigma_2b_spin}.}
\label{fig:diagrams}
\end{figure}

In this work, the correlation part of the self-energy is considered within the second Born (2B) approximation, diagrammatically represented in Fig.~\ref{fig:diagrams} and explicitly given by
\begin{align}\label{eq:sigma_2b_spin}
\Sigma_{ij}^{\sigma\sigma',<}(t,t') &= \sum_{\substack{\sigma_1\sigma_2\sigma_3 \\ \sigma_4\sigma_5\sigma_6}} \sum_{\substack{klm \\ nqp} }v_{iklm}^{\sigma\sigma_1\sigma_2\sigma_3} v_{nqpj}^{\sigma_4\sigma_5\sigma_6\sigma'}\times \\
&\left[G_{mn}^{\sigma_3\sigma_4,<}(t,t') G_{lq}^{\sigma_2\sigma_5,<}(t,t') G_{pk}^{\sigma_6\sigma_1,>}(t',t)\right. \nonumber \\
&\left.- G_{ln}^{\sigma_2\sigma_4,<}(t,t') G_{pk}^{\sigma_6\sigma_1,>}(t',t)G_{mq}^{\sigma_3\sigma_5,<}(t,t')\right], \nonumber
\end{align}
where all propagators implicitly depend on configuration $\mathcal{S}$ of the classical spin-texture, which also evolves in time.

Thus, according to Eqs.~\ref{eq:eom}-\ref{eq:sigma_2b_spin}, a GKBA for non-collinear spins can be achieved by interpreting all quantities as matrices in spin space, as e.g.
\begin{align}
\rho^{\sigma\sigma'}_{ij}  \rightarrow \rho_{i\sigma,j\sigma'} \equiv \begin{pmatrix} \rho_{\uparrow\uparrow}   & \rho_{\uparrow\downarrow} \\
                                          \rho_{\downarrow\uparrow} & \rho_{\downarrow\downarrow} \end{pmatrix}.
\end{align}
The notation can be made substantially lighter by using the spin-orbital index $m = 1,2,\ldots,2N$ and ordering the states according to $(1\uparrow,2\uparrow,\ldots,N\uparrow,1\downarrow,\ldots,N\downarrow)$. Since with a few exceptions (namely the construction of the Hartree-Fock Hamiltonian and the second Born self-energy) all quantities in the equations above are built solely using matrix multiplications, this entails very few changes to an existing GKBA implementation. In the super-index notation, Eqs.~\ref{eq:eom}-\ref{eq:g_ret} become identical to the usual spin-compensated case, and for the non-collinear case one uses
\begin{align}\label{eq:sigma_2b}
&[h_{HF}(t,\mathbfcal{S})]_{ij} = h_{ij}(t,\mathbfcal{S}) + \sum_{mn}\left[ v_{imnj} - v_{imjn} \right] \rho_{nm}(t), \nonumber \\
&\Sigma_{ij}^<(t,t') =\!\!\!
\sum_{klmnqp}v_{iklm}v_{nqpj} \left[G^<_{mn}(t,t') G^<_{lq}(t,t') G^>_{pk}(t',t)\right.\nonumber\\
&\left.- G^<_{ln}(t,t') G^>_{pk}(t',t)G^<_{mq}(t,t')\right]. 
\end{align}
As an example of the contracted notation, the Hubbard interaction $U\sum_i n_{i\uparrow}n_{i\downarrow}$ is written in the four-index, spin-orbital-index form as $v_{iklj} = U\delta_{ij}\delta_{kl}(\delta_{i+N,l}+\delta_{i,l+N})$. 

We note that within this formulation of the GKBA, it is possible to treat both spin polarized systems with different numbers of spin up and down electrons, as well as more general systems with non-collinear spin structure.

{\it Numerical implementation.-} As already discussed in the introduction (Sect.~\ref{Intro}), double-time NEGF codes for lattice models as developed e.g. in Ref.~\cite{Puig09} are not computationally convenient for problems of the size dealt with here.
For the implementation of our spin-dependent, quantum-classical mixed scheme, for i) the implementation of the 
time-diagonal NEGF approach, we used as starting point the
CHEERS code~\cite{Perfetto18-2}, and adapted it to a non-collinear spin description
in the presence of spin-orbit interactions. For ii) the implementation of the time-evolution of the classical spins within the
Ehrenfest approximation, we closely followed the procedure introduced in~\cite{Bostrom16,Balzer16}. Compared to the spin-compensated case, and depending on the nature of the interaction matrix elements and the type of many-body approximation for the
electronic self-energy, the non-collinear treatment presents an additional computational cost factor in the range $2^2 - 2^5$.


\section{Coupled system: Ground state}\label{Coupled}
After the initial state configuration of the classical spin lattice has been obtained, we need to find the initial state of the coupled spin-electron system. This is done by a self-consistent procedure minimizing the forces on the spins, and amounts to a damped time-evolution. During the minimization, the electronic system is treated at the Hartree-Fock level and is decoupled from the leads. 

For the damped spin dynamics, the forces are calculated via the equation of motion for the spin operators,
in the limit $\hbar \to 0$:
\begin{align}\label{eq:spin_force}
\frac{\partial {\bf S}_m}{\partial t} = &-2J\sum_n {\bf S}_n \times {\bf S}_m - {\bf h}(t)\times{\bf S}_m \\
&- 2D\sum_n \left[\hat{\bf e}_{mn}({\bf S}_m \cdot {\bf S}_n) - (\hat{\bf e}_{mn}\cdot{\bf S}_m) {\bf S}_n\right] \nonumber \\
&+ 4A_1 {\bf A}_m\times{\bf S}_m + A_2 {\bf B}_m\times{\bf S}_m \nonumber \\
&- \sum_{i\sigma\sigma'}J'_{im} \rho_{i\sigma,i\sigma'}({\boldsymbol \sigma}_{\sigma\sigma'}\times {\bf S}_m), \nonumber
\end{align}
The first five terms are due to spin-spin interactions, where the contributions proportional to the anisotropies are given in terms of the vectors ${\bf A}_m = ([S_m^x]^3,[S_m^y]^3,[S_m^z]^3)$ and ${\bf B}_m = (S_{m+\hat{\bf x}},S_{m+\hat{\bf y}},0)$.
The last term in Eq.~\ref{eq:spin_force} describes the interaction of the spin-texture with the electrons spins in the wire.
It corresponds to performing the Ehrenfest approximation for Eq.~\ref{eq:interaction}, by taking the average of the electronic spins over the instantaneous electronic state of the wire~\footnote{This way to proceed mirrors the prescription employed in molecular dynamics simulations to describe the semi-classical interaction between electrons and nuclei or spin, introduced within the framework of NEGF in~\cite{Bostrom16,Balzer16}. Recently, the semiclassical approach has also been considered to study of spin-electron interactions in a single classical spin model~\cite{Stahl17}}.

\begin{figure}
 \includegraphics[width=\columnwidth]{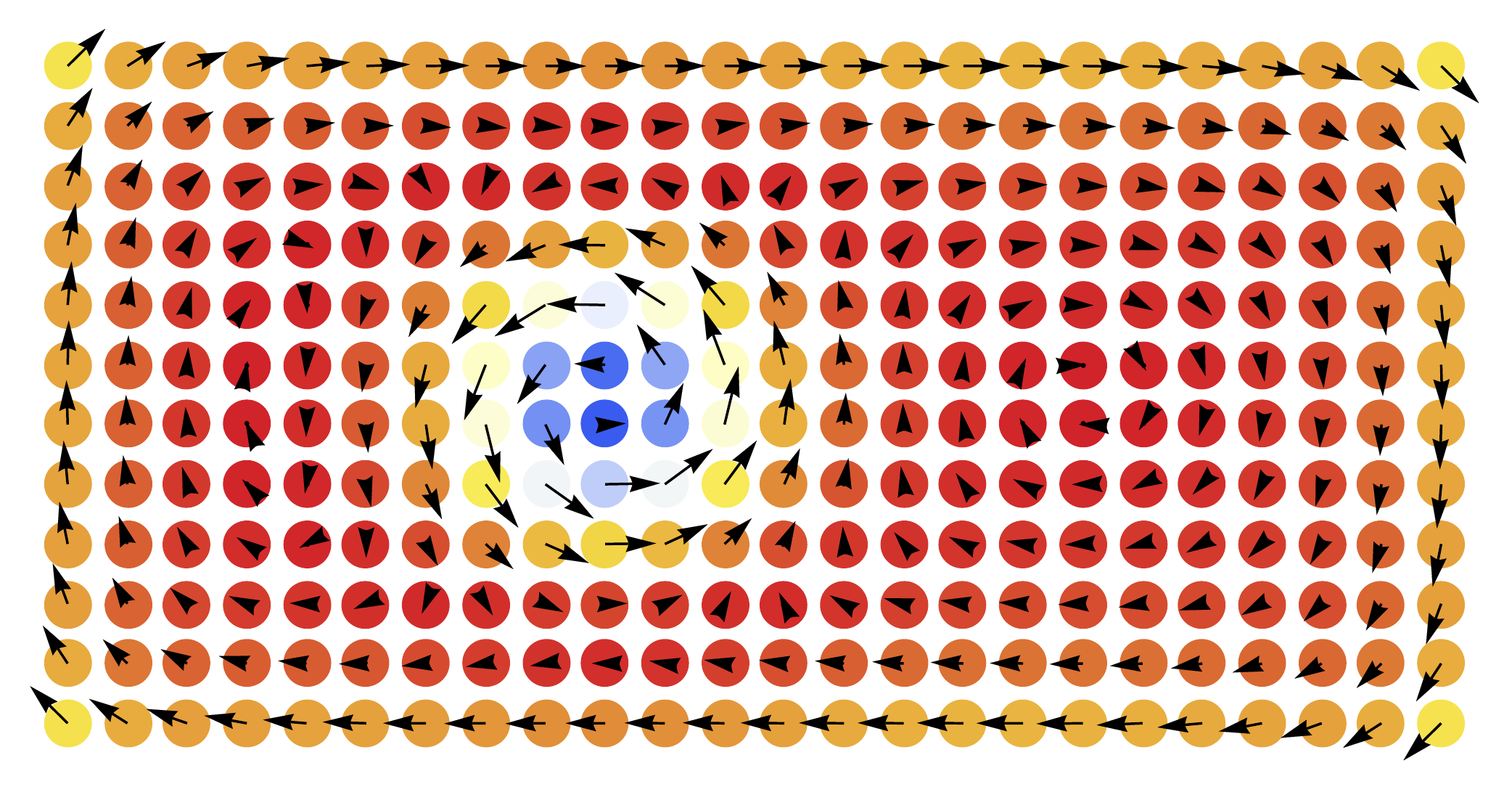}
 \includegraphics[width=\columnwidth]{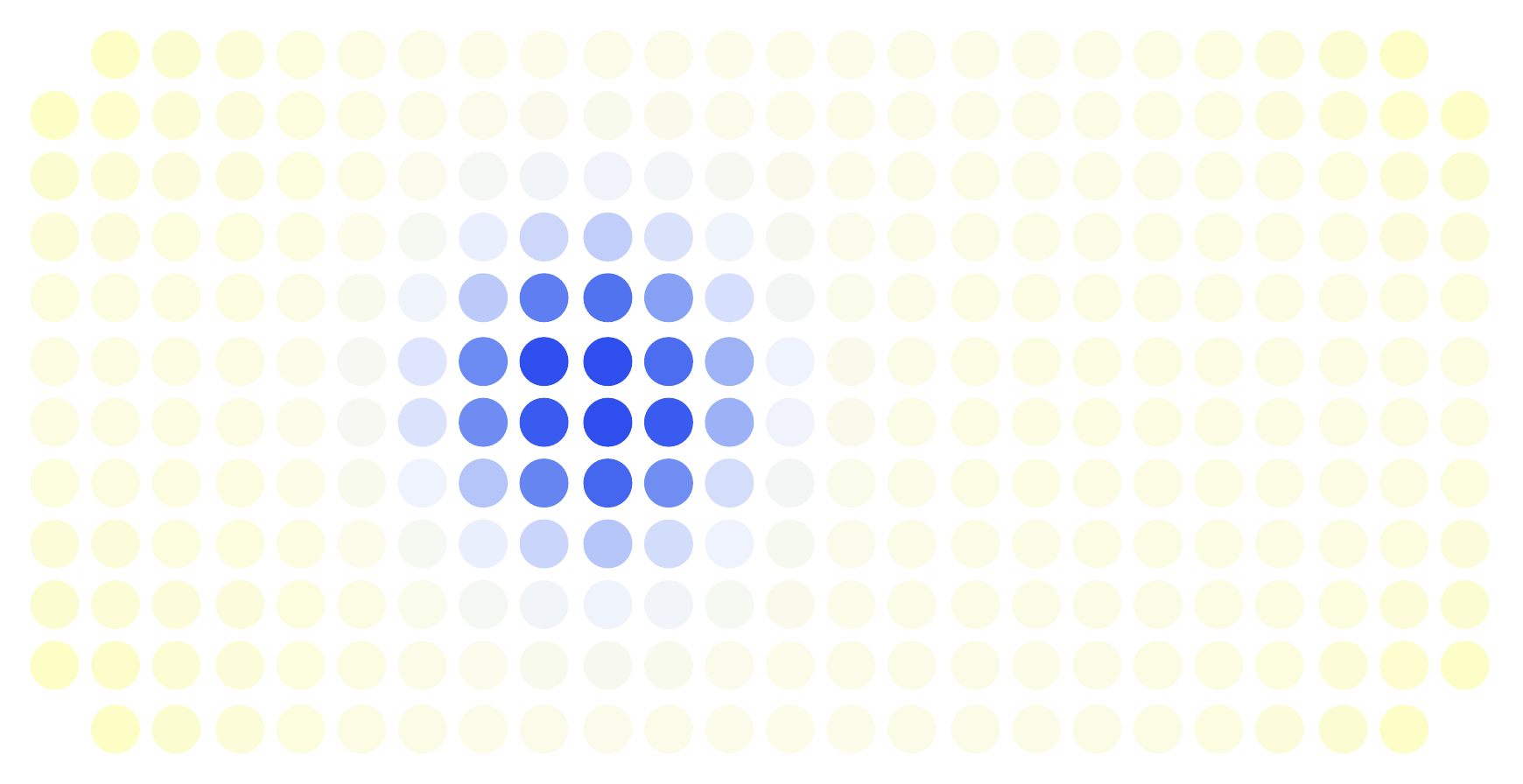}
 \includegraphics[width=\columnwidth]{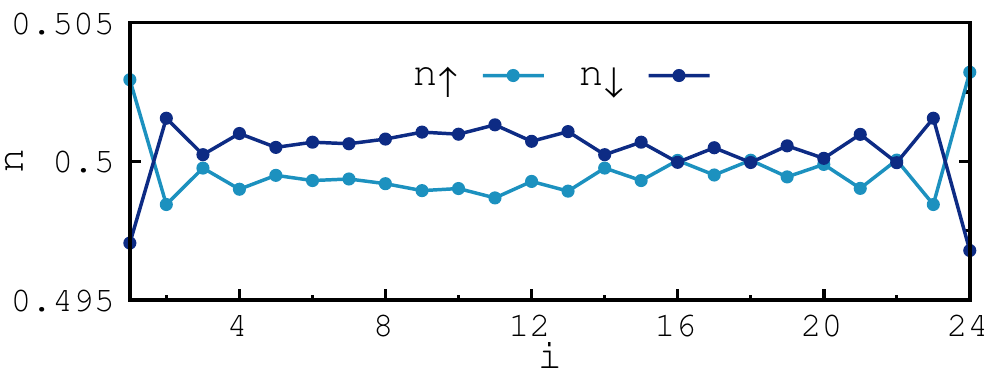}
 \caption{Spin configuration (top) and skyrmion density (center) in the initial state of a coupled spin-electron system. The bottom panel shows the density profile of spin up and down electrons across the wire. Color coding specifications are the same as in Fig.~\ref{fig:model}.}
 \label{fig:initial_state_long}
\end{figure}

While determining the initial state of the coupled system (and only then), we add for numerical convenience an additional term to Eq.~\ref{eq:spin_force}, namely a Gilbert-like damping force ${\bf F}_G$ explicitly given by
\begin{align}
{\bf F}_{m,G} = a_G {\bf S}_m \times \left(\frac{\partial}{\partial t}{\bf S}_m \right),
\end{align}
This acts as a dissipative term that quenches the spin oscillations, and permits to attain the ground state faster~\cite{Han17}. The damped self-consistent dynamics to obtain the ground state is performed using a predictor-corrector scheme. The procedure can be summarized into the following steps:
\begin{enumerate}
\item ~Calculate the electronic potential due to the spins, Eq.~\ref{eq:interaction}.
\item ~Calculate the spin forces according to Eq.~\ref{eq:spin_force}.
\item ~Find the Hartree-Fock ground state of the electrons.
\item ~Update the spin configuration.
\end{enumerate}
The above is repeated until the maximal force on any of the spins is below a given threshold value. 

The procedure described is not restricted to starting from the spin ground state, and in fact, we could take any initial state of the spin system, couple it to the electrons, and then relax the system to find a stationary state. This is a point worth stressing since, most experiments on single skyrmions are performed in the ferromagnetic parameter regime with isolated skyrmion excitations~\cite{Romming13,Woo16}.

To study skyrmion motion we consider a rectangular lattice with $24\times 12$ sites, coupled to a $24$ site electronic wire. The initial state is obtained by first computing the spin ground state of $12\times 12$ lattice, using the spin parameters $J = 0.5$, $D = 1/\sqrt{6}$, $h = 0.3$, $A_1 = 0.2$, $A_2 = 0$ and $a_G = 1$, and the electronic parameters $t = 1$, $t_{so} = 0$, $\epsilon_i = 0$ and $U = 0$. The spin-electron coupling strength is taken as $J' = 0.5$. We then extend the lattice along the $x$-direction, and relax the system in presence of a slightly increased magnetic field $h = 0.5$. This value of the magnetic field corresponds to a spin ground state that is ferromagnetic, so that the stationary state shown in Fig.~\ref{fig:initial_state_long} corresponds to a single skyrmion excitation. We see that the density profile of the electronic ground state is slightly asymmetric with respect to the center, which however is no surprise, since also the spin structure has this type of asymmetry.


\section{Coupled system: Time evolution}\label{Timeevo}
The time evolution of the coupled spin-electron system is obtained by propagating the GKBA equation of motion for the electrons (Eq.~\ref{eq:eom}), together with the equation of motion for the spins (Eq.~\ref{eq:spin_force}), using a predictor-corrector scheme for both spins and the electrons. We consider leads with complete spin polarization, and to model spin- and charge-currents we therefore connect the central system to four leads: Two left leads with spin up and down respectively, and two right leads with spin up and down. We consider reservoirs close to the wide band limit by taking $t' = 9$ and $t_l = t_r = 0.2$, and fix the chemical potential at $\mu = 0$. We have checked that the results presented below are robust against an increase of the value of $t'$ (i.e. $t'>$9); thus, for all practical purposes, the wide band limit appears to have been numerically attained in our calculations.

\begin{figure}
 \includegraphics[width=\columnwidth]{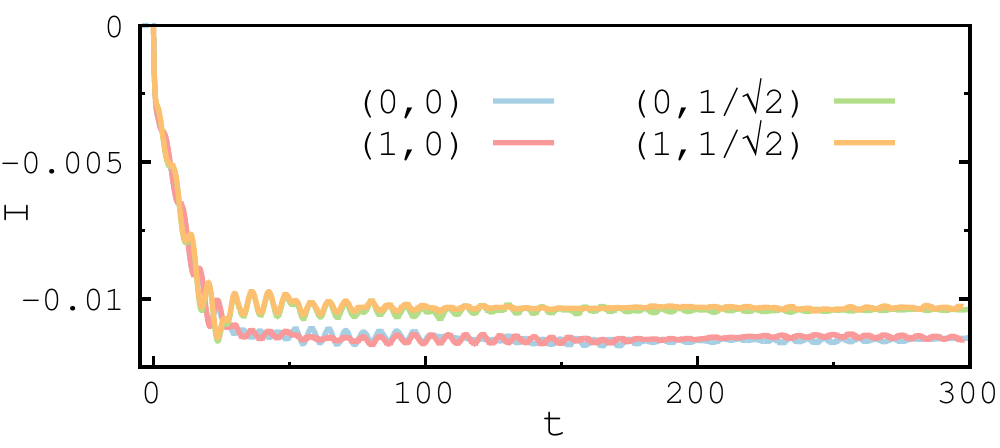}
 \caption{Time-dependent current $I = I_\uparrow + I_\downarrow$ from the left leads into the wire, after a sudden switch-on of the bias $u_{l\sigma}(t) = \theta(t)$ and $u_{r\sigma}(t) = -\theta(t)$. The different curves represent the values of the parameters $(U,t_{so})$ indicated in the figure.}
 \label{fig:current}
\end{figure}

Starting from the initial state of the $24\times 12$ lattice discussed above, the coupling to the leads is slowly switched on between times $t =-50$ and $t = 0$. At time $t = 0$ a spin symmetric bias given by $u_{l\sigma}(t) = \theta(t)$ and $u_{r\sigma}(t) = -\theta(t)$ is suddenly switched on, which generates a charge current density of about $I = I_\uparrow + I_\downarrow = 0.01$ flowing through the system, as seen in Fig.~\ref{fig:current}. From the figure, it is also evident that the overall features of the current are insensitive to the exact parameters of the wire, both when varying (in a not too wide range) the interaction $U$ and the spin-orbit hopping $t_{so}$.

The current exerts a force on the spin texture through the so-called spin-transfer torque, which induces a motion of the skyrmion pattern. It has been found in several studies that the skyrmion tends to follow the electronic current~\cite{Iwasaki13,Sampaio13}. This effect is typically modeled by assuming a large spin-electron coupling, which allows to derive an effective equation for the spins that includes electronic off-diagonal spin couplings up to second order~\cite{Volovik87}. In this equation, the effects of electronic motion is included only through an external current density $I$, which is usually taken constant or as a static solution of the macroscopic Maxwell's equations.

Here we also consider skyrmion motion, but we adopt a microscopic description of {\it both} electrons and spins. While being computationally more expensive, this approach permits to explicitly address issues not easily accessible otherwise, e.g. how electron-electron interactions or disorder affect currents, or how time dependent currents
permit to manipulate skyrmion dynamics. Furthermore, it provides a firm conceptual and practical ground to benchmark  treatments where the electrons are considered only implicitly.

\begin{figure}
 \includegraphics[width=\columnwidth]{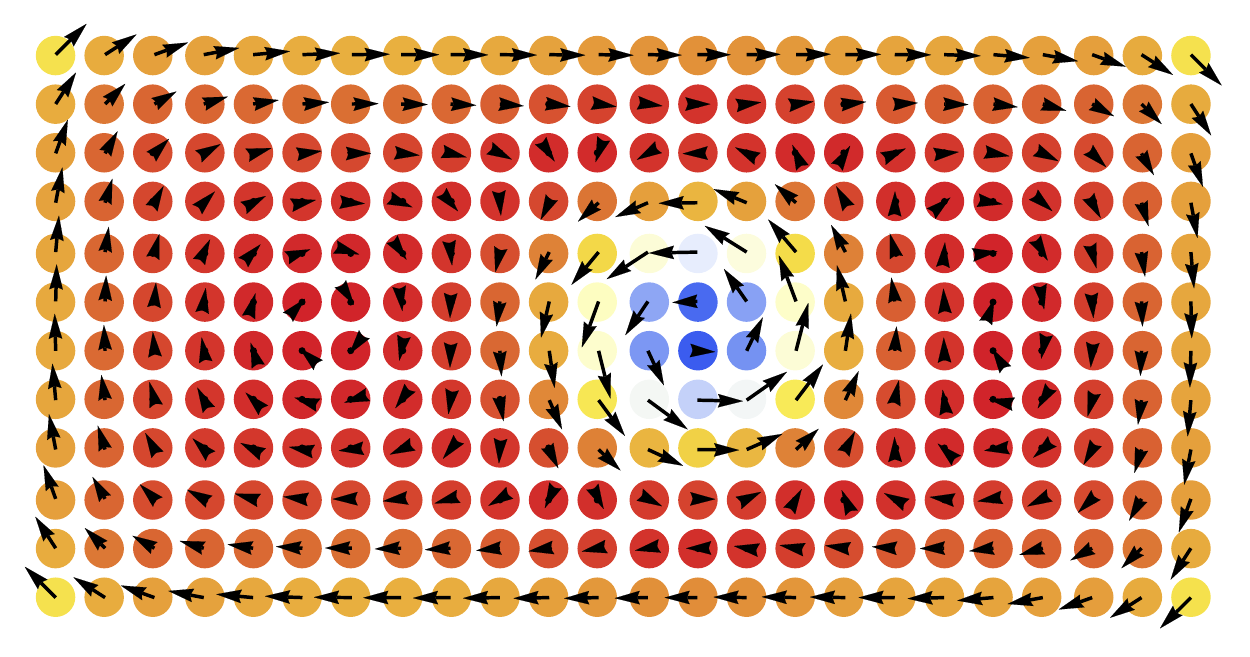}
 \includegraphics[width=\columnwidth]{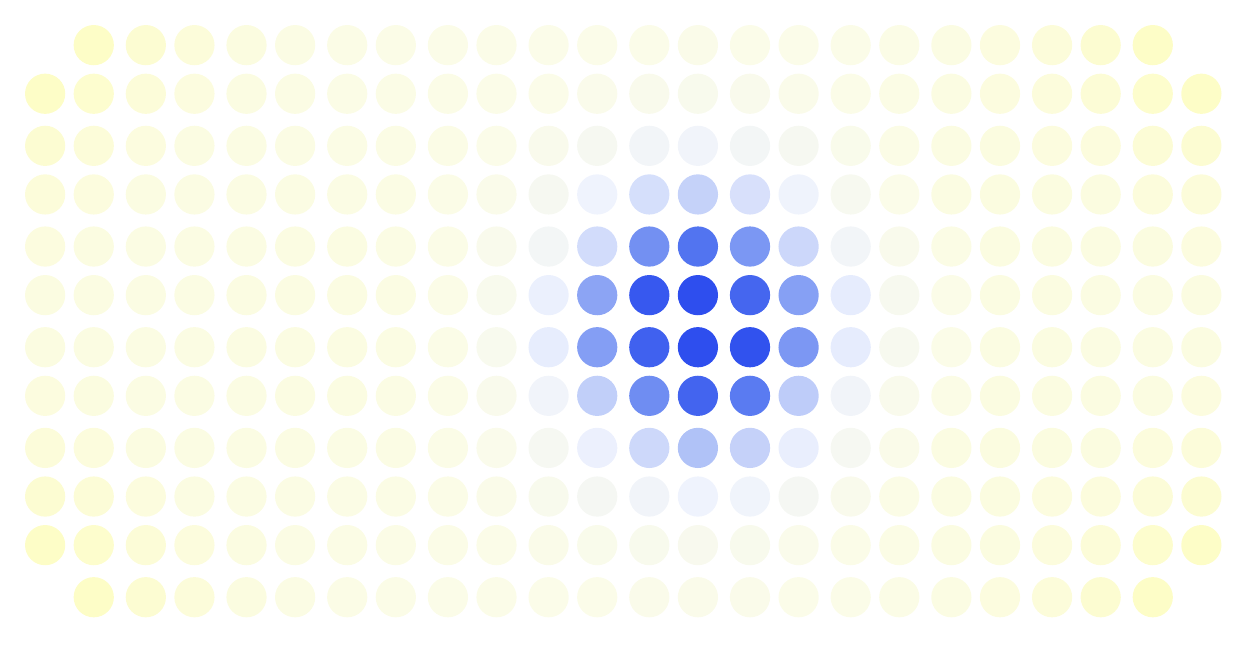}
 \caption{Spin configuration (top) and skyrmion density (bottom) at time $t = 200$, for a wire with $U = t_{so} = 0$. Color coding specifications are the same as in Fig.~\ref{fig:model}.}
 \label{fig:skyrmion_motion}
\end{figure}

In Fig.~\ref{fig:skyrmion_motion} we show the spin configuration at time $t = 200$ for $U = t_{so} = 0$, where we see that the current in the wire induces an almost rigid motion of the skyrmion. As in previous studies we find that the skyrmion moves in the direction of the current (cf. the initial configuration in Fig.~\ref{fig:initial_state_long}). However in our approach, this motion arises from a microscopic description of both electrons and spins, and without any assumptions about the relative strength of the parameters. We have checked that these results are insensitive to changes in the interaction $U$ and spin-orbit coupling $t_{so}$, as well as to moderate variations of the bias. We have also tried using spin-polarized currents, taking the bias to be $u_{l\uparrow} = u_{r\downarrow} = \theta(t)$ and  $u_{l\downarrow} = u_{r\uparrow}= -\theta(t)$, which gives a current $I = I_\uparrow - I_\downarrow \approx 0.01$. The results are in close agreement with those using the spin symmetric bias, and are therefore not shown explicitly here.

For a more quantitative characterisation of the motion of the skyrmion in Fig.~\ref{fig:skyrmion_motion}, we resort to three 
distinct, time-dependent indicators: i) For the motion as a whole, we define center of mass coordinate of the spin structure as
\begin{align}
 {\bf R}(t) = \frac{\sum_m |\mathbb{\varrho}^\mathcal{SK}_m(t)|{\bf R}_m}
 {\sum_m |\mathbb{\varrho}^\mathcal{SK}_m(t)|}. 
\end{align}
Since $\mathbb{\varrho}^\mathcal{SK}$ can be negative, we use a modified definition in terms of the absolute value of $\mathbb{\varrho}^\mathcal{SK}_m(t)$. Additionally, ii) to address the deformation and spread of the skyrmion, we use a modified inverse participation ratio (IPR), defined through
\begin{align}
 \chi(t) \equiv \frac{  \sum_m |\mathbb{\varrho}^\mathcal{SK}_m(t)|^2}{\big[\sum_m |\mathbb{\varrho}^\mathcal{SK}_m(t)|\big]^2}. 
\end{align}
This quantity (again defined in terms of $|\mathbb{\varrho}^\mathcal{SK}_m |$) lies in the range $\chi \in [1/N,1]$, where $N$ is the number of spins. A value $\chi = 1$ corresponds to a perfectly localized density (to a single spin), while the value $\chi = 1/N$ corresponds to complete delocalization. Finally, iii) to address the stability of the skyrmion, we look at the standard definition of topological charge,
\begin{align}
 Q(t) = \sum_m \mathbb{\varrho}^\mathcal{SK}_m(t).
\end{align}
In a continuum description this quantity can only take on integer values, but in a lattice description it is no longer perfectly quantized.

\begin{figure}
 \includegraphics[width=\columnwidth]{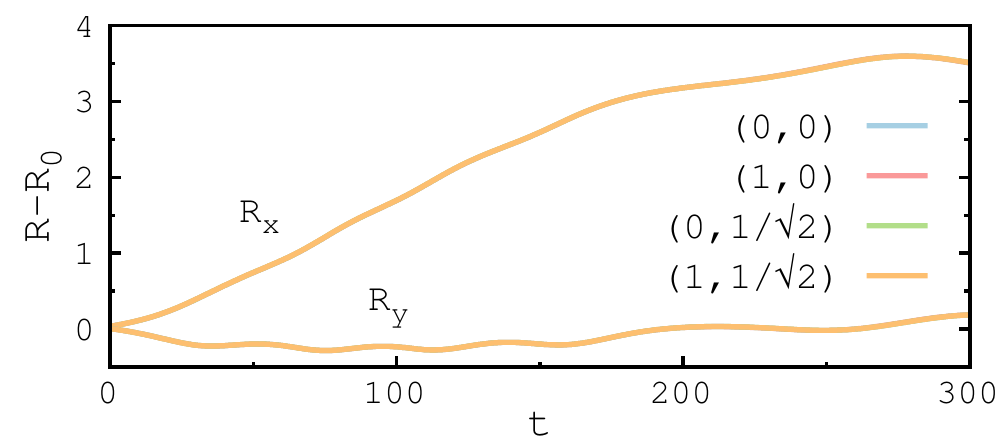}
 \includegraphics[width=\columnwidth]{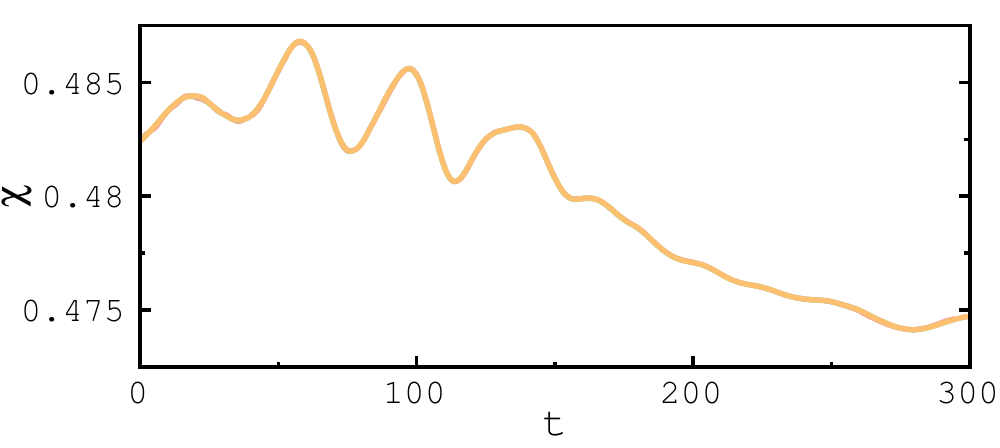}
 \includegraphics[width=\columnwidth]{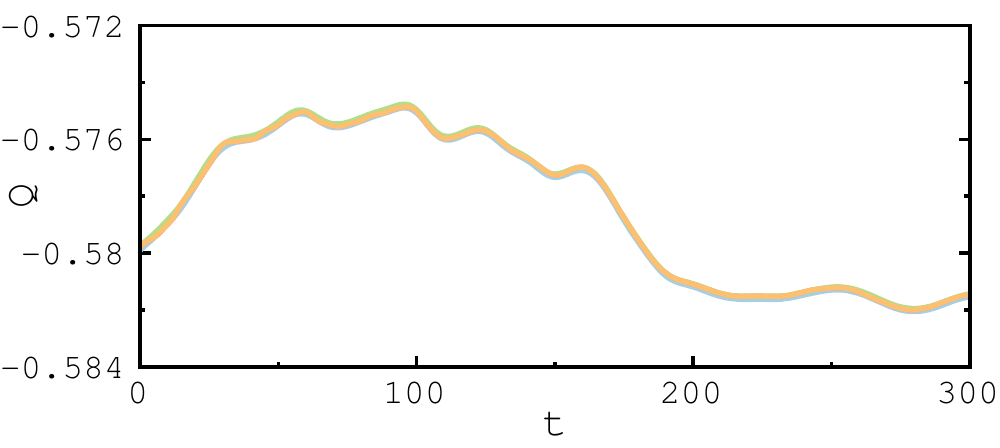}
 \caption{Top panel: Time evolution of the skyrmion center of mass ${\bf R}$, measured from the initial state value ${\bf R}_0\equiv {\bf R}(t=0)$. Middle panel:  inverse participation ratio $\chi$ as a function of time. Bottom panel: time-dependent topological charge $Q$. In each panels, the different curves represent the values of the parameters $(U,t_{so})$ indicated in the top panel.}
 \label{fig:center}
\end{figure}

In Fig.~\ref{fig:center} we show the time-evolution of the center of mass ${\bf R}$, the IPR $\chi$ and the topological charge $Q$. These results clearly show that the skyrmion moves along the positive $x$-direction, while its position along the $y$-direction remains approximately constant. We also see that the IPR only changes slightly while the topological charge remains close to constant during the time evolution. Taken together, the results of Fig.~\ref{fig:center} provide a clear indication of rigid skyr-mion motion induced by the electronic current. In addition, this result seems highly robust against changes in the electronic parameters.

\section{Conclusions}\label{conc}
We have introduced a theoretical description of the motion of
classical magnetic skyrmions as induced by time-dependent spin-resolved currents. 
In our approach, spin-carrying electronic currents and skyrmion textures are
treated microscopically and explicitly, via a mixed quantum-classical scheme
where the spin texture producing the skyrmions is described classically, and the electrons in the spin-current-carrying 
wire are described with nonequilibrium Green's functions in a time-diagonal formulation.
As an illustration, we used our mixed quantum-classical scheme to study a single Bloch-type skyrmion on a ferromagnetic
2D square lattice, showing how, as times goes by, the skyrmion can be dragged along a current carrying wire.

As well as evidence from time snapshots of the skyr-mion texture, this outcome stems from results for the time evolution of the skyrmion center of mass, the skyrmion inverse participation ratio, and the skyrmion charge. We have also made simple and limited explorations about the role of electron-electron and
spin-orbit interaction in the wire, and found that these seem to have
a minor effect on skyrmion dynamics. However, we have at present too little evidence to
make this a general statement, and we plan to perform further investigation is this respect.

Besides electron-electron and spin-orbit interactions, notions such disorder, time-dependent driving,
optimally controlled skyrmion steering, multiple wires, and skyrmion/ wire circuitry are in principle all within the scope of our
approach. It is also the case that our mixed quantum-classical scheme
is not confined to the study of skyrmion phenomena; rather, it is expected to be relevant in all those situations where classical-spin
systems interact with time-dependent quantum spin currents. Furthermore, results obtained within our methodology can potentially provide benchmarks 
to other methods, where the role of the electrons is considered only implicitly, subsumed in a description based 
directly on currents.  The investigation of these different aspects is left to future work.


\section*{Acknowledgements}~ We wish to thank Daniel Cabra and
Angel Rubio for fruitful discussions, and E. Perfetto and G. Stefanucci for providing us with
an early version of the CHEERS code. We also gratefully acknowledge 
support from Craafoordska Stiftelsen, grant No. 2017010.

\bibliographystyle{pss}
\bibliography{references}

\end{document}